\documentclass[aps,twocolumn,pra,superscriptaddress,amsmath,showpacs,tightenlines,pdflatex,longbibliography]{revtex4-1}
\usepackage{amssymb}
\usepackage{amsmath}
\usepackage{dcolumn}
\usepackage{graphicx}
\usepackage{mathrsfs}
\usepackage{appendix}
\usepackage{graphicx}
\usepackage{booktabs}
\usepackage{color}

\newcommand{\ket}[1]{\mbox{$|#1\rangle$}}
\newcommand{\bra}[1]{\mbox{$\langle#1|$}}

\usepackage{url}
\usepackage[colorlinks]{hyperref}
\hypersetup{%
    plainpages=true,
    breaklinks=true,
    hypertexnames=false,
    pageanchor=true,
    colorlinks=true,
    linkcolor={blue},
    citecolor={red},
    urlcolor={blue},
    anchorcolor={black}
}

\begin{document}

\title{Multi-output microwave single-photon source
using superconducting circuits with longitudinal and transverse
couplings}

\author{Xin Wang}
\affiliation{Institute of Quantum Optics and Quantum Information,
School of Science, Xi'an Jiaotong University, Xi'an 710049, China}
\affiliation{CEMS, RIKEN, Wako-shi, Saitama 351-0198, Japan}

\author{Adam Miranowicz}
\affiliation{CEMS, RIKEN, Wako-shi, Saitama 351-0198, Japan}
\affiliation{Faculty of Physics, Adam Mickiewicz University,
61-614 Pozna\'n, Poland}

\author{Hong-Rong Li}
\affiliation{Institute of Quantum Optics and Quantum Information,
School of Science, Xi'an Jiaotong University, Xi'an 710049, China}

\author{Franco Nori}
\affiliation{CEMS, RIKEN, Wako-shi, Saitama 351-0198, Japan}
\affiliation{Physics Department, The University of Michigan, Ann
Arbor, Michigan 48109-1040, USA}

\date{\today}

\begin{abstract}
Single-photon devices at microwave frequencies are important for
applications in quantum information processing and communication
in the microwave regime. In this work, we describe a proposal of
a multi-output single-photon device. We consider two
superconducting resonators coupled to a gap-tunable qubit via
both its longitudinal and transverse degrees of freedom. Thus,
this qubit-resonator coupling differs from the coupling in standard circuit
quantum-electrodynamic systems described by the Jaynes-Cummings model. We
demonstrate that an effective quadratic coupling between one of
the normal modes and the qubit can be induced, and this induced
second-order nonlinearity is much larger than that for
conventional Kerr-type systems exhibiting photon blockade.
Assuming that a coupled normal mode is resonantly driven, we
observe that the output fields from the resonators exhibit strong
sub-Poissonian photon-number statistics and photon antibunching.
Contrary to previous studies on resonant photon blockade, the
first-excited state of our device is a pure single-photon Fock
state rather than a polariton state, i.e., a highly hybridized
qubit-photon state. In addition, it is found that the optical
state truncation caused by the strong qubit-induced nonlinearity
can lead to an entanglement between the two resonators, even in
their steady state under the Markov approximation.
\end{abstract}

\pacs{42.50.Ar, 42.50.Pq, 85.25.-j} \maketitle

\section{Introduction}
In quantum information, the generation, distribution, and storage
of quantum information at the single-photon level are of great
importance~\cite{Kiraz04,Kimble08,Brien09}. Therefore,
single-photon sources of non-classical light states are
needed~\cite{Faraon08,He13}. In some cases, we can reduce the
power of a laser or maser source to avoid large probabilities of a
multi-photon output. However, the field might be of an extremely
low intensity. Photon sources differ not only by their frequencies
and polarizations, but also by the statistical properties of the
emitted photons~\cite{Walls07}. Photons from a coherent source are
still classical, while in proposals of security for the quantum
cryptography~\cite{Scarani09} the sources of single-photons
exhibiting strong antibunching and sub-Poissonian statistics can
help to avoid eavesdropping on an encode message.

To increase the output rate of such sources of non-classical
fields, one requires some form of nonlinearity. For example,
single-photon manipulation can be realized via photon blockade
(see Refs.~\cite{Tian92,Leonski94,Miran96,Imamoglu97,
Verger06,Miran13,Majumdar13,Liu14a,Xu14} and references therein),
in which the nonlinearity prevents more than a single excitation
being exited in a cavity: Only when the first photon has left the
cavity can another identical photon be reexcited. Photon blockade
originates from the anharmonic energy-level structure in nonlinear
systems. It has been predicted and demonstrated experimentally in
platforms such as optical cavities with a trapped
atom~\cite{Birnbaum05}, integrated photonic crystal cavities with
a quantum dot~\cite{Faraon11,Majumdar12}, or  microwave
transmission-line resonators (superconducting ``cavities'') with a
single superconducting artificial atom~\cite{Hoffman11,Lang11}.
Recently, photon blockade and closely related phonon blockade were
predicted in optomechanical systems (see,
e.g.,~\cite{Liu10,Didier11,Rabl11,Liao13,Xu13,Wang15}). In the
earlier studies, the observation of conventional photon blockade
requires large nonlinearities with respect to the decay rate of
the system. More recently, it was found that strong
entanglement~\cite{Leonski94,Miran06} and strong photon
antibunching~\cite{Liew10,Stassi2015} can be generated via
destructive quantum interference in coupled nonlinear oscillators:
Transition paths for multi-excitations cancel each other and, as a
result, the population of the two-photon state is effectively
suppressed. This underlying mechanism is called ``unconventional
photon blockade" and further research has been devoted to it in
various kinds of
systems~\cite{Bamba11,Ferretti13,Lemonde14,Kyriienko14,Zhou15,Tang15}.
It is worth mentioning that the idea of using photon blockade as a
single-photon turnstile device was suggested already in the first
theoretical works on this effect~\cite{Leonski94,Imamoglu97}.

The standard single-photon blockade has also been generalized to
multiphoton blockade, which is also referred to as photon
tunneling. These multiphoton effects have not only been described
theoretically (see,
e.g.,~\cite{Miran13,Hovsepyan14,Miran14qre,Miran14,Wang15,Miran16}
and references therein), but even demonstrated
experimentally~\cite{Smolyaninov02,
Faraon08,Majumdar12,Schuster08,Kubanek08}. Such multiphoton
effects are often discussed in the context of optical state
truncation (for a review see Ref.~\cite{Miran01,Leonski01}). Here
we focus on the standard single-photon blockade, although we
also show that multiphoton processes can also be induced in our
system.

Recent developments on superconducting quantum devices provide
versatile artificial quantum systems for quantum communication and
information
processing~\cite{You05,Buluta11,You2011,Xiang13,DiCarlo10,Lucero12}.
For example, methods for microwave-photon detection based on
superconducting quantum circuits have been demonstrated in
Refs.~\cite{Romero09,Peropadre11,Chen11,Sathyamoorthy14,Inomata16}.
Moreover, schemes for measuring photon statistics in the microwave
regime have also been proposed both in theoretical and
experimental studies~\cite{Lang13,Virally16}. All these progresses
have laid a solid foundation for applications at the single-photon
level based on superconducting circuits. Therefore, efficient and
well-performed single-photon devices in the microwave regime are
very important, and have been studied. Resonant photon blockade
has been observed in a quantum circuit composed of a
superconducting qubit and a transmission-line
resonator~\cite{Lang11}. Moreover, Ref.~\cite{Ridolfo12},
discussed the effect of ultrastrong coupling on photon blockade in
circuit quantum electrodynamics (QED) systems. All these schemes
require the qubit and resonator to be resonant.
 In another approach \cite{Hoffman11}, the
dispersive microwave photon blockade was predicted due to the
$\chi ^{(3)}$ nonlinearity (about $\sim $1~MHz), which can be
induced by a qubit. The sub-Poissonian photon statistics and
photon antibunching were also predicted in such systems.

Here we introduce another mechanism to obtain microwave-photon
blockade via the effective quadratic coupling in a
circuit-QED-based system. Our scheme is composed of two resonators
and a single qubit. Different from standard circuit-QED systems
with Jaynes-Cummings coupling, our system is based on both
longitudinal and transverse couplings. We demonstrate that, in
principle, arbitrary multiphoton processes can be induced in our
system. In particular, we obtain the effective Hamiltonian for the
quadratic coupling between one supermode and the qubit. As opposed
to the resonant photon blockade, the first excitation of this
system is a bare single-photon state, rather than hybridized with
the qubit excited state (i.e., a polariton state), which might
provide higher tolerance to imperfections in
experiments~\cite{Hoffman11}. The second-order nonlinear coupling
strength can be of tens of MHz under current experiment
approaches, which is much stronger than the induced $\chi ^{(3)}$
nonlinearity in superconducting systems \cite{Liu14a,Miran16}.
With a stronger nonlinearity, we can consider resonators with
higher-photon escape rates and apply stronger coherent drive
fields for the two resonators, and the single-photon output fields
can be of much higher intensities. By modeling the quantum input
and output fields from channels of independent resonators and
joint channels of two resonators, we find that all the three
output fields are antibunched and sub-Poissonian in photon-number
statistics, so our proposal can serve as an efficient
single-microwave photon source with multi-output channels.

The organization of this paper is as follows: In Sec.~I, we
describe the layout of the model consisting of a qubit and two
superconducting resonators, and then we analytically derive the
Hamiltonian for multi-photon processes in the two resonators. In
Sec.~II, we demonstrate how to employ the effective quadratic
coupling between the qubit and the resonators to achieve
single-photon blockade in the two resonators. After that, we find
that it is possible to apply our system as a microwave single
photon device with multi-output channels. In Sec.~III, we show our
numerical results. In particular, we analyze nonclassical
photon-number correlations and give a phase-space description of
the single-mode (single-resonator) states generated via photon
blockade. The last section presents our final discussions and
conclusions.

\section{Model}
\subsection{Circuit layout and Hamiltonian}
\begin{figure}[tbp]
\centering \includegraphics[width=8.6cm]{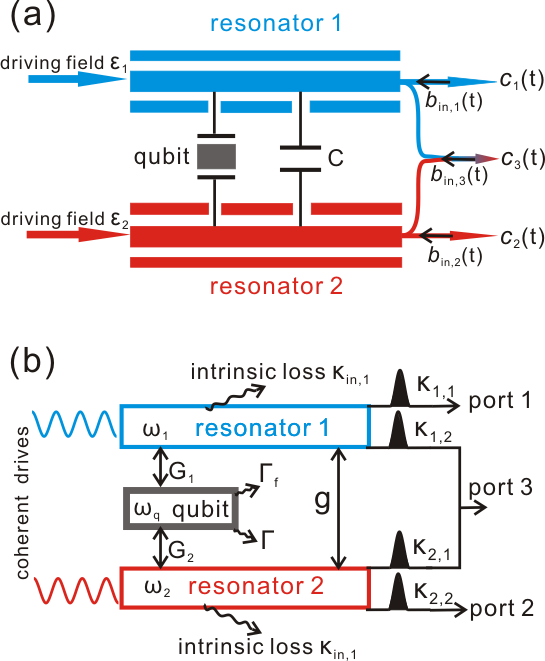} \caption{(Color
online) (a) Schematic circuit layout and (b) the couplings and
dissipative  channels of our proposal. A gap-tunable qubit
(e.g., a flux or charge qubit) couples with two
superconducting (e.g., transmission-line) resonators with
strengths $G_{i}$ for $i=1,2$. The eigenfrequencies for the
$i$th resonator and the qubit are $\omega_{i}$ and $\omega_{q}$,
respectively. A capacitor $C\ $is used to directly mediate the two
resonators and results in a coupling strength $g$. In the
left-hand side, two coherent microwave drives, with strengths
$\protect\epsilon _{1}$ and $\epsilon _{2}$, are applied to the
resonators 1 and 2, respectively. In the right-hand side, the
single photon output microwave photons are collected from ports 1,
2, and 3. Ports 1 and 2 are semi-infinite transmission lines
connected to resonators 1 and 2. This results in photon escape
rates $\protect\kappa _{1,1} $ and $\protect\kappa _{2,2}$,
respectively. Port 3 is the joint output transmission line from
resonators 1 and 2, with photon escape rates $\protect\kappa
_{1,2}$ and $\protect\kappa _{2,1}$. We assume that the input
fields $b_{\text{in},i}$ for these three ports are all independent
vacuum noises. Beside escaping into the transmission lines, the
photons in the two resonators can also dissipate into the
environment with intrinsic rates $\protect\kappa _{\rm{in},i}$.
For the qubit, the decay (dephasing) rate is $\Gamma$
($\Gamma_{f}$). } \label{fig01}
\end{figure}
As schematically shown in Fig.~\ref{fig01}, we consider a
gap-tunable superconducting artificial atom, such as a charge or
flux qubit, coupled with two superconducting resonators of
frequencies $\omega _{1}$ and $\omega
_{2}$~\cite{Strauch10,Li12,Zhao15,Garziano15}. Moreover, we assume
that the coupling between the resonators is directly mediated via a
capacitance $C$~\cite{Underwood12}. The Hamiltonian can be written
as
\begin{figure}[tbp]
\centering
\includegraphics[width=7.4cm]{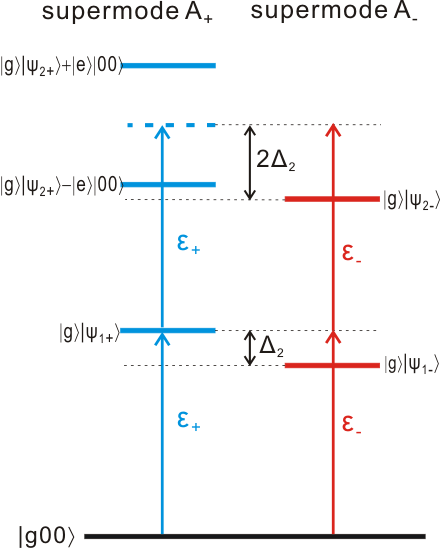}
\caption{(Color online) The lowest energy levels for the
Hamiltonian in Eq.~(\protect\ref{eq13}). The supermode $A_{+}$
couples with the qubit with quadratic form, while the supermode
$A_{-}$ decouples from the qubit. The frequency difference between
these two supermodes is $\Delta _{2}$. When $\Delta _{2}=0$, these
two modes are degenerate. The effective drives for the supermodes
$A_{+}$ and $A_{-}$ are  $\protect\epsilon _{+}$ and
$\protect\epsilon _{-}$, respectively, as shown in Table~I.}
\label{fig02}
\end{figure}
\begin{eqnarray}
\bar{H}_{0} &=&\frac{1}{2}\omega \bar{\sigma}_{z}+\frac{1}{2}\Delta \bar{\sigma}_{x}+\sum_{i=1,2}\omega _{i}a_{i}^{\dag }a_{i} \notag \\
&&+g(a_{1}^{\dag }a_{2}+a_{2}^{\dag
}a_{1})+\bar{\sigma}_{z}\sum_{i=1,2}G_{i}(a_{i}^{\dag }+a_{i}),
\label{eq1}
\end{eqnarray}
where $a_{i}$ ($a_{i}^{\dag}$) denotes the annihilation (creation)
operator for the $i$th resonator, $g$ is the coupling constant
between the two resonators due to the hopping capacitor $C$, and
$G_{i}$ is the coupling strength between the $i$th resonator and
the qubit. Here we assume that $g\ll G_{i}$, which justifies the
use of the rotating-wave approximation (RWA) in Eq.~(\ref{eq1}).
The Pauli spin operators $\bar{\sigma}_{z}$ and $\bar{\sigma}_{x}$
are defined in the basis of the two quantum states of the qubit,
and $\omega$ and $\Delta$ are the energy bias and tunable qubit
gap, respectively. In experiments, both $\omega$ and $\Delta$ can
be controlled independently via external parameters. For example,
in a flux qubit~\cite{Mooij99,You07,You20081,Paauw09}, $\Delta $
can be tuned by applying an external flux though the
superconducting quantum interference loop, while $\omega $ can be
adjusted by controlling the flux though the qubit loop.

Note that we assume that the inter-resonator coupling $g$ is much
smaller than the qubit-resonator couplings $G_{i}$. Moreover,
these couplings can be strong but not ultrastrong, to justify the
application of the RWA. However, the RWA is not valid in the
ultrastrong coupling regime, where at least one of the couplings
$G_{i}$ is comparable or larger than the corresponding resonator
frequency $\omega_i$. In such a case, due to the counter-rotating
terms, photon blockade effects are usually significantly changed
compared to the standard blockade under the RWA (see
Refs.~\cite{Ridolfo12,Boite16}). For example, Ref.~\cite{Boite16}
showed that multiple antibunching-to-bunching transitions can be
observed when increasing the resonator-qubit coupling strength in
the standard (i.e., transverse) Rabi model. These transitions lead
to the vanishing and reappearance of photon blockade due to the
presence of the counter-rotating terms, which modify the
nonlinearity of the energy spectrum and can cause two-photon
cascade decays. We expect that similar effects can be observed in
our model if the inter-resonator and qubit-resonator coupling
constants are increased.

In the qubit basis, we can write the Hamiltonian in
Eq.~(\ref{eq1}) as
\begin{eqnarray}
H_{0} &=&\frac{1}{2}\omega _{q}\sigma _{z}+\sum_{i=1,2}\omega
_{i}a_{i}^{\dag }a_{i}+g(a_{1}^{\dag }a_{2}+a_{2}^{\dag }a_{1})  \notag \\
&&+\sum_{i=1,2}\left[G_{x,i}\sigma _{x}(a_{i}^{\dag
}+a_{i})+G_{z,i}\sigma _{z}(a_{i}^{\dag }+a_{i})\right],
\label{eq2}
\end{eqnarray}
where the coupling constants $G_{x,i}=-G_{i}\sin \theta $ and
$G_{z,i}=G_{i}\cos \theta $ describe, respectively, the transverse
and longitudinal couplings between the qubit and the resonators,
with $\tan \theta =\Delta /\omega$, and $\omega _{q}=\sqrt{\omega
^{2}+\Delta ^{2}}$ is the transformed qubit eigenfrequency.

In a typical picture of a circuit-QED system, the interaction
between cavities and artificial atoms is transverse, which can be
simplified to Jaynes-Cummings-type models under the
rotating-wave approximation. Another alternative layout for
circuit-QED is based on the longitudinal qubit-cavity
interaction~\cite{Wang09,Liu14,Didier15,Wang16}. The Hamiltonian
in Eq.~(\ref{eq2}) describes a qubit with both transverse and
longitudinal couplings to the resonators. In such an artificial
system, multiphoton Rabi oscillations between a single resonator
and a qubit have been predicted in Ref.~\cite{Garziano15}. In the
following discussions, by considering a more general case with two
resonators coupled with a qubit, we will analytically obtain the
effective Hamiltonians for arbitrary multiphoton processes between
the two resonators and the qubit.

We apply two coherent driving fields for the two resonators with
strengths $\epsilon _{1}$ and $\epsilon _{2}$, respectively, as
shown in Fig.~\ref{fig01}. Under the rotating-wave approximation,
the corresponding driving Hamiltonian is
\begin{equation}
H_{d}=\sum_{i=1,2}(\epsilon _{i}a_{i}^{\dag }e^{-i\omega
_{d,i}t}+\epsilon _{i}^{\ast }a_{i}e^{i\omega _{d,i}t}),
\label{eq3}
\end{equation}
and the total Hamiltonian for the system can be expressed as
\begin{equation}
H_{s}=H_{0}+H_{d}.  \label{eq4}
\end{equation}
The two driving fields might have a phase difference $\theta$. By
assuming that $\epsilon _{1}=|\epsilon _{1}|e^{-i\theta /2}$ and
$\epsilon _{2}=|\epsilon _{2}|e^{i\theta /2},$ we will in
Sec.~IV.B show that both the relative phase $\theta $ and the
drive strength $|\epsilon _{i}|$ have significant effects on the
photon distribution statistics of the output fields.
\begin{table*}[tbp]
{\normalsize \renewcommand\arraystretch{2.3}
\begin{tabular}{>{\hfil}p{1.7in}<{\hfil}>{\hfil}p{2.4in}
<{\hfil}>{\hfil}p{2.1in}<{\hfil}}
\hline\hline & Supermode $A_{+}$ & Supermode $A_{-}$ \\
\hline Eigenfrequencies & $\Omega _{+}=\omega
_{1}+\frac{g}{\beta},$ $\Omega _{+}^{^{\prime}}=\Omega _{+}
-\frac{4G_{x}^{2}}{3\Omega _{+}} $ & $\Omega
_{-}=\omega _{2}-\frac{g}{\beta}$ \\
Driving strengths & $\epsilon _{+}=\frac{\beta \epsilon
_{1}+\epsilon _{2}}{\sqrt{1+\beta ^{2}}}$ & $\epsilon
_{-}=\frac{\epsilon _{1}-\beta \epsilon
_{2}}{\sqrt{1+\beta ^{2}}}$ \\
First-excited states & $|\psi _{1+}\rangle =\frac{\beta |10\rangle
+|01\rangle }{\sqrt{1+\beta ^{2}}}$ & $|\psi _{1-}\rangle
=\frac{|10\rangle
-\beta |01\rangle }{\sqrt{1+\beta ^{2}}}$ \\
Second-excited states & $|\psi _{2+}\rangle =\frac{\beta
^{2}|20\rangle +\sqrt{2}\beta |11\rangle +|02\rangle }{1+\beta
^{2}}$ & $|\psi _{2-}\rangle =\frac{|20\rangle -\sqrt{2}\beta
|11\rangle +\beta ^{2}|02\rangle }{1+\beta ^{2}}$ \\
Effective nonlinear coupling & couples with the qubit & decouples
from the qubit \\ \hline\hline
\end{tabular}}
\caption{The parameters and eigenstates of the two supermodes
$A_{+}$ and $A_{-}$, according to the Hamiltonian in
Eq.~(\protect\ref{eq13}).} \label{table1}
\end{table*}

\subsection{Multiphoton processes}
To explicitly demonstrate multiphoton processes between the qubit and the
two resonators, we first introduce the two supermodes
via their annihilation operators:
\begin{subequations}
\begin{align}
A_{+}& =\frac{G_{1}a_{1}+G_{2}a_{2}}{\overline{G}},  \label{eq5a} \\
A_{-}& =\frac{G_{2}a_{1}-G_{1}a_{2}}{\overline{G}},  \label{eq5b}
\end{align}where $\overline{G}=\sqrt{G_{1}^2+G_{2}^2}=G_1\sqrt{1+\beta^2}$, and the commutation
relation between $A_{i}$ and $A_{j}^{\dag }$ is
[$A_{i},A_{j}^{\dag }]=\delta _{ij}.$ We define $\beta
=G_{1}/G_{2}$ as the ratio of the coupling strengths. The detuning
between the resonator fundamental frequencies should satisfy the
relation
\end{subequations}
\begin{equation}
\omega _{1}-\omega _{2}=g(\beta ^{2}-1)/\beta.  \label{eq6}
\end{equation}

Assuming that the two drives are of the same frequency
$\omega _{d,i}=\omega _{d}$, we express $H_{s}$ in Eq.~(\ref{eq4})
in terms of $A_{+} $ and $A_{-}$ as follows:
\begin{widetext}
\begin{equation}
H_{s}=\frac{1}{2}\omega _{q}\sigma _{z}+\sum_{i=\pm }\Omega
_{i}A_{i}^{\dag }A_{i}+G_{z}\sigma _{z}(A_{+}^{\dag
}+A_{+})+G_{x}\sigma _{x}(A_{+}^{\dag }+A_{+})+\sum_{i=\pm
}\epsilon _{i}(A_{i}^{\dag }e^{-i\omega
_{d}t}+\text{H.c.}),  \label{eq7}
\end{equation}\end{widetext}
where the renormalized eigenfrequencies $\Omega _{\pm}$ and the
driving strengths $\epsilon _{\pm}$ for the supermodes $A_{\pm}$
are presented in Table~\uppercase\expandafter{\romannumeral1}, and
\begin{subequations}
\begin{align}
G_{z}& =\overline{G}\cos \theta,  \label{eq8a} \\
G_{x}& =-\overline{G} \sin \theta  \label{eq8b}
\end{align}are the longitudinal and transverse coupling strengths between the
qubit and the supermode $A_{+}$, respectively. From
Eq.~(\ref{eq7}), we find that the supermode $A_{-}$ decouples from
the qubit. Let us apply the frame rotated by the unitary polariton
transformation $\exp [-\lambda \sigma _{z}(A_{+}^{\dag }-A_{+})]$,
with $\lambda =G_{z}/\Omega _{+}$~\cite{Zhao15,Wilson04}, and use
the commutation relation
\end{subequations}
\begin{equation*}
\lbrack A_{+},f(A_{+},A_{+}^{\dag })]=\frac{\partial
f(A_{+},A_{+}^{\dag })}{\partial A_{+}},
\end{equation*}
where $f(A_{+},A_{+}^{\dag})$ can be expanded in a power series of
the operators $A_{+}$ and $A_{+}^{\dag }$. Then the total
Hamiltonian becomes
\begin{widetext}
\begin{equation}
{H}_{s}=\frac{1}{2}\omega _{q}\sigma _{z}+\sum_{i=\pm }\Omega
_{i}A_{i}^{\dag }A_{i}+G_{x}\left\{\sigma _{+}\left[A_{+}^{\dag
}f(\lambda )+f(\lambda
)A_{+}\right]+\text{H.c.}\right\}+\sum_{i=\pm }\epsilon
_{i}(A_{i}^{\dag }e^{-i\omega
_{d}t}+\text{H.c.})-2\lambda \epsilon _{+}\sigma _{z}\cos (\omega
_{d}t),  \label{eq9}
\end{equation}\end{widetext}
where $f(\lambda)=\exp [2\lambda (A_{+}^{\dag }-A_{+})]$. For weak
driving strengths $\epsilon _{1,2}$ and a small Lamb-Dicke
parameter $\lambda \ll 1$~\cite{Leibfried03}, $2\lambda \epsilon
_{+}$ is a much smaller parameter. Therefore the last term in
$\tilde{H}_{s}$ can be neglected.

Let us now show how to realize multi-photon processes by setting
$\omega _{q}\simeq n\Omega _{+}$, with $n$ being the order of the
photon-qubit transitions. By expanding the third terms in
Eq.~(\ref{eq9}) in terms of the small parameter $\lambda $, and keeping only
the resonant terms, we obtain the corresponding Hamiltonian for the
$n$-photon processes,
\begin{eqnarray}
H_{n}&=G_{x}\sum_{m=1}^{\infty }\left[B_{1}(m,n)\sigma
_{+}A_{+}^{\dag
m}A_{+}^{m+n}+\text{H.c.}\right]  \notag \\
&+G_{x}\sum_{m=0}^{\infty }\left[B_{2}(m,n)\sigma _{+}A_{+}^{\dag
m}A_{+}^{m+n}+\text{H.c.}\right]\text{,}  \label{eq10}
\end{eqnarray}where the coefficients $B_{i}(m,n)$ are expressed
as
\begin{subequations}
\begin{align}
B_{1}(m,n)& =e^{-2\lambda ^{2}}\frac{(-1)^{m+n}(2\lambda )^{2m+n-1}}{(m-1)!(m+n)!},  \label{eq11a} \\
B_{2}(m,n)& =e^{-2\lambda^{2}}\frac{(-1)^{m+n-1}(2\lambda)^{2m+n-1}}{m!(m+n-1)!}.
\label{eq11b}
\end{align}
\end{subequations}
From Eq.~(\ref{eq10}), we conclude that, in principle, arbitrary
multi-photon processes between the qubit and one supermode of the
two resonators can be induced in this circuit-QED system. However,
for a small parameter $\lambda $, the rates of $n$-photon
transitions, which are determined by $B_{1}(m,n)$ and
$B_{2}(m,n)$, decrease rapidly with increasing $m $ and $n$; so
higher-order photon-qubit transitions have slower rates and,
therefore, can be overwhelmed by the rapid oscillation terms and
decoherence channels.

In experiments, the interaction in a circuit-QED system can easily
enter into the strong-coupling
regime~\cite{Chiorescu04,Johansson06,Abdumalikov08}. For two
superconducting resonators oscillating at frequency $\omega
_{i}/(2\pi)$=2.5~GHz with $G_{i} $=$0.06\omega _{i}$, and by
setting $\theta =\pi/4$, the rates for the two-photon ($n=2$) and
three-photon transitions ($n=3) $ between the qubit and the
supermode $A_{+}$ are $\Theta _{2}/(2\pi) \simeq $18~MHz and
$\Theta _{3}/(2\pi) \simeq $1.1~MHz, respectively.

Here we assume that the qubit should be operated around its
optimal point (but not exactly at this point), so the
dephasing noise of the qubit is the dominant decoherence channel.
As reported in Ref.~\cite{Yoshihara06}, for a flux qubit operated
around the optimal point (the flux bias is $\Phi _{b}\sim
$1$\times $10$^{-3}\Phi _{0}$, with $\Phi _{0}$ being the flux
quantum), the dephasing rate was measured about 6~$\mu $s$^{-1}$
(the corresponding dephasing rate $\Gamma _{f}/(2\pi) \simeq$
1~MHz). The quality factor $Q$ of a superconducting resonator can
easily exceed 10$^{4}$~\cite{Megrant12} (i.e., the decay rate
$\gamma /(2\pi) \simeq $0.25~MHz). Thus, the rate for the
two-photon (three-photon) transitions exceeds (is comparable to)
all the decoherence rates in current experimental implementations,
and it is possible to observe quantum coherent phenomena due to
these multiphoton processes.

\bigskip
\section{Analytical results}
\subsection{Photon blockade in two resonators}
In this part, we will demonstrate the single-photon blockade in
the two resonators, which can be induced by the two-photon
processes. By setting $\omega _{q}\simeq 2\Omega _{+}\gg G_{x}$
and neglecting the last term, we expand Eq.~(\ref{eq9}) to
first order in $\lambda$, and obtain
\begin{widetext}
\begin{equation}
{H}_{s}\cong\frac{1}{2}\omega _{q}\sigma _{z}+\sum_{i=\pm}\Omega
_{i}A_{i}^{\dag }A_{i}+G_{x}\sigma _{x}(A_{+}^{\dag
}+A_{+})+2\lambda G_{x}\left[\sigma _{+}(A_{+}^{\dag
2}-A_{+}^{2})+\text{H.c.}\right]+\sum_{i=\pm }\epsilon
_{i}(A_{i}^{\dag }e^{-i\omega _{d}t}+\text{H.c.}).
\label{eq12}
\end{equation}\end{widetext}

The effective Hamiltonian for the third term can be expressed as
$4G_{x}^{2}/\left( 3\Omega _{+}\right) \sigma _{z}A_{+}^{\dag
}A_{+}$, which can be viewed as the dispersive coupling between
the qubit and the supermode $A_{+}$~\cite{Puri16}. In this paper,
we find that the qubit remains effectively in its ground
state, so this term will only renormalize the eigenfrequency of
the supermode $A_{+}$ to $\Omega _{+}^{^{\prime }}=\Omega
_{+}-4G_{x}^{2}/\left( 3\Omega _{+}\right) $. Assuming $\omega
_{q}=2\Omega _{+}^{^{\prime}}$, and performing the unitary
transformation
\begin{equation}
U=\exp \left\{-i\omega _{d}\sigma _{z}t-i\sum_{i=\pm }\omega
_{d}A_{i}^{\dag }A_{i}t\right\},  \label{unitarytrans}
\end{equation}
we obtain the following time-independent Hamiltonian by neglecting
the fast-oscillating terms in $H_{s}$
\begin{eqnarray}
H_{\text{eff}}&=&\frac{1}{2}\Delta _{+}\sigma _{z}+\sum_{i=\pm
}\Delta
_{i}A_{i}^{\dag}A_{i}  \notag \\
&+&\Theta \left(\sigma _{+}A_{+}^{2}+\sigma _{-}A_{+}^{\dag
2}\right)+\sum_{i=\pm }\epsilon _{i}(A_{i}^{\dag }+A_{i}),
\label{eq13}
\end{eqnarray}
where $\Theta =-2\lambda G_{x}$, $\Delta _{+}=\Omega
_{+}^{^{\prime }}-\omega _{d}$ is the frequency detuning between
the supermode $A_{+}$ and the drive field, and $\Delta _{-}=\Delta
_{+}+\Delta _{2}$ with
\begin{equation}
\Delta _{2}=\frac{4G_{x}^{2}}{ 3\Omega _{+}} -\frac{g(1+\beta
^{2})}{\beta}  \label{eq14}
\end{equation}
is the frequency difference between these two supermodes, which
can be obtained from the parameters in Eq.~(\ref{eq6}) and
Table~\uppercase\expandafter{\romannumeral1}. The third term in
Eq.~(\ref{eq13}) describes the quadratic coupling between the
supermode $A_{+}$ and the qubit, while the supermode $A_{-} $
decouples from the qubit. Moreover, the supermode $A_{+} $
($A_{-}$) is driven with strength $\epsilon _{+}$ ($\epsilon
_{-}$) and detuning $\Delta _{+}$ ($\Delta _{-}$).

As shown in Table~\uppercase\expandafter{\romannumeral1}, the
ground state of the system is $|g\rangle \otimes |\psi _{0}\rangle
=|g\rangle|00\rangle$, and the first-excited states for the
supermodes $A_{+}$ and $A_{-}$ are the single-photon entangled
states
\begin{equation}
|\psi _{1\pm}\rangle =\frac{G_{1} |10\rangle \pm
G_{2}|01\rangle}{\sqrt{G_{1}^{2}+G_{2}^{2}}}.  \label{eq15}
\end{equation}
Without the nonlinear coupling of the resonators with the qubit,
the second-excited states for the supermodes $A_{+}$ and $A_{-}$
become $|\psi _{2+}\rangle $ and $|\psi _{2-}\rangle $,
respectively, which are defined by
\begin{equation}
|\psi _{2\pm}\rangle =\frac{G_{1}^{2} |10\rangle \pm
G_{1}G_{2}|10\rangle +G_{2}^{2}|01\rangle }{G_{1}^{2}+G_{2}^{2}}.
\label{state2}
\end{equation}
However, due to the effective nonlinear coupling, the second
excited states for supermode $A_{+}$ are the two dressed (as
marked by the subscript $d$) states
\begin{equation}
|\Psi _{d,\pm}\rangle =\frac{|g\rangle|\psi _{2+}\rangle
\pm|e\rangle|00\rangle}{\sqrt{2}}  \label{dstates}
\end{equation}
with energy splitting $2\sqrt{2}\Theta$, as shown in
Fig.~\ref{fig02}. As a consequence, the energy levels of supermode
$A_{+}$ become anharmonic. It should be noted that $\epsilon _{\pm
}$ can conveniently be adjusted by changing the pumping strengths
$\epsilon _{1}$ and $\epsilon _{2}$, as presented in
Table~\uppercase\expandafter{\romannumeral1}. Hence, under the
conditions: $\Delta _{-}\gg $ $\epsilon _{-}$, or
$\epsilon _{-}\simeq 0$, the supermode $A_{-}$ cannot be
driven effectively. Meanwhile, if the supermode $A_{+}$ are
resonantly driven with strength $\epsilon _{+}$, the state $|\psi
_{1+}\rangle $ will be occupied, and the first photon can enter
into the two resonators. However, the two-photon state $|\psi
_{2+}\rangle $ can hardly be excited due to the non-existence of
available states. Thus, for the two resonators, the two-photon
states $|20\rangle $, $|02\rangle,$ and $|11\rangle $ will be of
extremely low probabilities. Similar to the case in
Refs.~\cite{Leonski04,Miran06}, the Hilbert space of this
composite system is only spanned by the vacuum and single-photon
states. These two resonators behave as a qubit with the ground and
excited states being $|\psi _{0}\rangle =|00\rangle $ and $|\psi
_{1+}\rangle $, respectively.

\subsection{Input-output relations for the three ports}
We consider the input and output ports as sketched in
Fig.~\ref{fig01}. At the outer edges, each resonator is
capacitively coupled to two semi-infinite transmission
lines~\cite{Underwood12}. By combining one transmission line of
each resonator as port 3~\cite{Flayac13}, we achieve three input
and output ports. Here we discuss the first-order correlation features of the output field from these three
channels. The second-order correlation functions will be discussed
in Sec.~IV.C.

The corresponding boson operators for the input and output modes
of the $i$th port are denoted as $b_{\text{in},i}$ and
$c_{i}$, respectively. According to the input-output relations,
the input, output, and intra-resonator fields are linked through
the boundary conditions~\cite{Flayac13,Collett84,Gardiner85},
\begin{subequations}
\begin{align}
c_{1}& =b_{\text{in},1}+\sqrt{\kappa _{1,1}}a_{1},  \label{eq16a} \\
c_{2}& =b_{\text{in},2}+\sqrt{\kappa _{2,1}}a_{2},  \label{eq16b} \\
c_{3}& =b_{\text{in},3}+\sqrt{\kappa _{1,2}}a_{1}+\sqrt{\kappa
_{2.2}}a_{2}, \label{eq16c}
\end{align}
where $\kappa _{i,j}$ is the photon escape rate from the resonator
$i$ to its $j$th line~\cite{Underwood12,Flayac13}. With the
intrinsic loss rate $\kappa _{\text{in,}i\text{ }}$of the
resonator $i$, the total loss rate for this resonator can be
expressed as $\gamma _{i}=\kappa _{\text{in,}i\text{ }}+$ $\kappa
_{i,1}+\kappa _{i,2}.$ Without loss of generality, we assume
that the decay rates of all the channels for each resonator are
the same, i.e., $\kappa _{\text{in,}i\text{ }}=\kappa
_{i,j}=\frac{\gamma _{i}}{3}$ for $i,j=1,2$. Moreover, the input
fields $b_{\text{in},i}$ of the three ports are all independent
quantum vacuum noises, and satisfy the Markov correlation
relations:
\end{subequations}
\begin{equation}
\langle b_{\text{in},i}(t)b_{\text{in},j}^{\dag}(t^{^{\prime
}})\rangle =\delta (t-t^{^{\prime }})\delta _{ij},  \label{eq17}
\end{equation}so all the normally-ordered cross correlations between
the intra-resonator and input field are zero, and the correlations
of output fields at each port can be expressed only with the
resonator operators. The average output photon numbers collected
through the three ports, which are proportional to the first-order
correlation functions with the zero-time delay, can be expressed
as
\begin{subequations}
\begin{align}
N_{1}& =\langle c_{1}^{\dag }c_{1}\rangle =\frac{\gamma
_{1}}{3}\langle
a_{1}^{\dag }a_{1}\rangle ,  \label{eq18a} \\
N_{2}& =\langle c_{2}^{\dag }c_{2}\rangle =\frac{\gamma
_{2}}{3}\langle
a_{2}^{\dag }a_{2}\rangle ,  \label{eq18b} \\
N_{3}& =\langle c_{3}^{\dag }c_{3}\rangle =\frac{\gamma
_{1}\langle a_{1}^{\dag }a_{1}\rangle +\gamma _{2}\langle
a_{2}^{\dag }a_{2}\rangle }{3}
\notag \\
&\quad +\frac{\sqrt{\gamma _{1}\gamma _{2}}(\langle a_{1}^{\dag
}a_{2}\rangle+\langle a_{2}^{\dag }a_{1}\rangle )}{3}.
\label{eq18c}
\end{align}
\end{subequations}
We will apply these input-output relations, in particular, in
Sec.~IV.C.

\begin{figure}[tbp]
\centering \includegraphics[width=8.6cm]{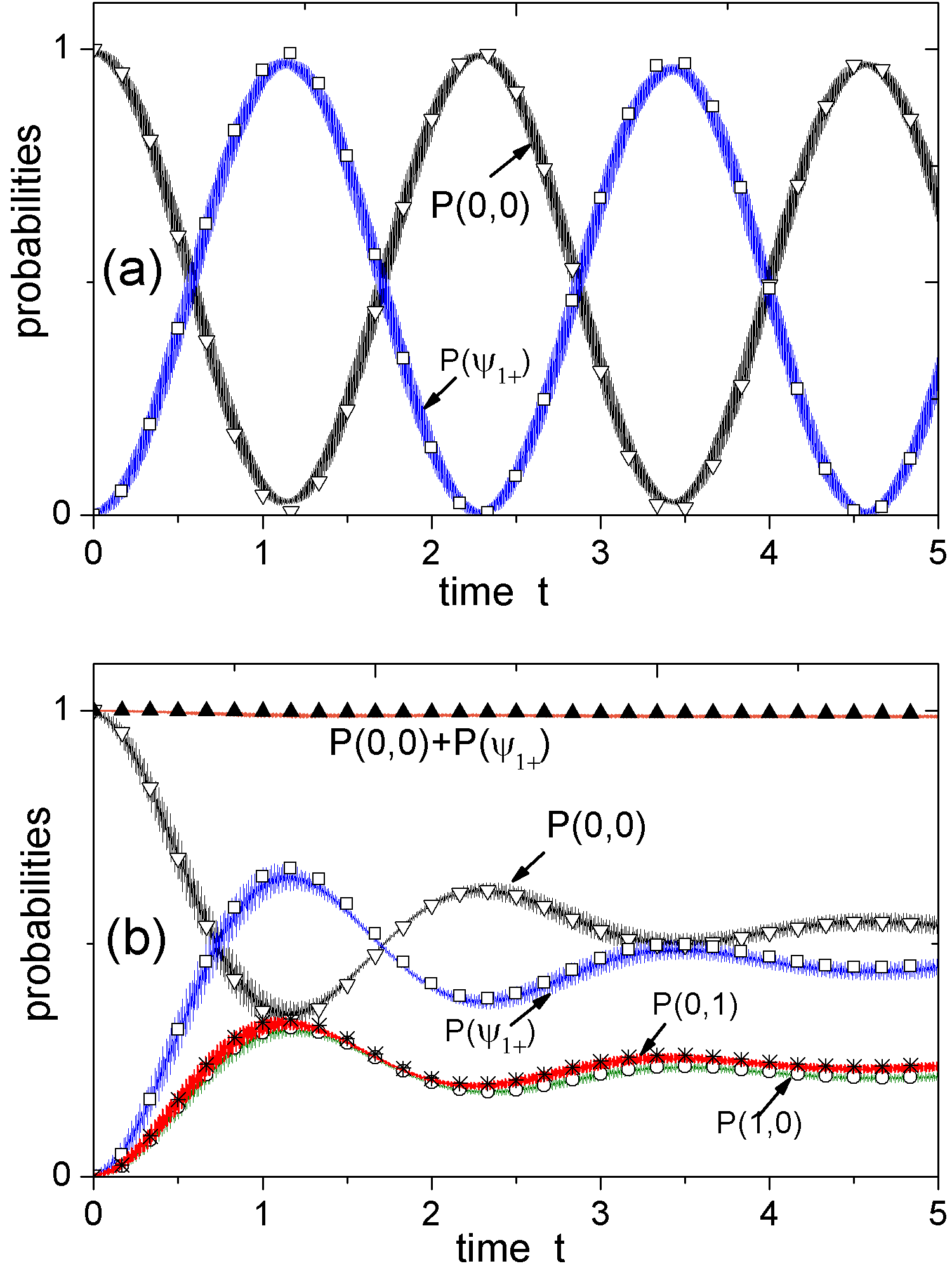} \caption{(Color
online) Time evolutions of the probabilities for the system
described by $H=H_{s}$ (shown with solid curves) and
$H=H_{\text{eff}}$ (marked with symbols) in the (a) nondissipative
and (b) dissipative cases. The initial state is $|0,0\rangle
|g\rangle$. (a) Without considering any decay channels, the
time-evolution of the probabilities $P(0,0)$ and $P(\protect\psi
_{1+})$ exhibit the Rabi oscillations between the states
$|0,0\rangle$ and $P(\protect\psi _{1+})$. (b) The decay of the
probabilities assuming the decoherent rates $\Gamma=\Gamma
_{f}/2=\protect\gamma _{1}=\protect\gamma _{2}=1$. We find that
the sum of $P(0,0)$ and $P(\protect\psi _{1+})$ is almost
equal to 1 for all the evolution times. Thus, this sum can be
considered as a fidelity measure of optical state truncation
resulting in photon blockade. Here we consider that the two modes
are degenerate, i.e., $\Delta _{2}=0$.} \label{fig03}
\end{figure}

\section{Numerical results}
\subsection{Time-dependent solutions of the master equation}

In this section, we numerically demonstrate that
single-photon blockade can occur in our system even assuming
amplitude and phase damping as described by the master equation.
Numerical computations of the time-evolution solution of the
master equation were performed using the Python package
QuTiP~\cite{Johansson12qutip,Johansson13}.

\begin{figure}[tbp]
\centering \includegraphics[width=8.5cm]{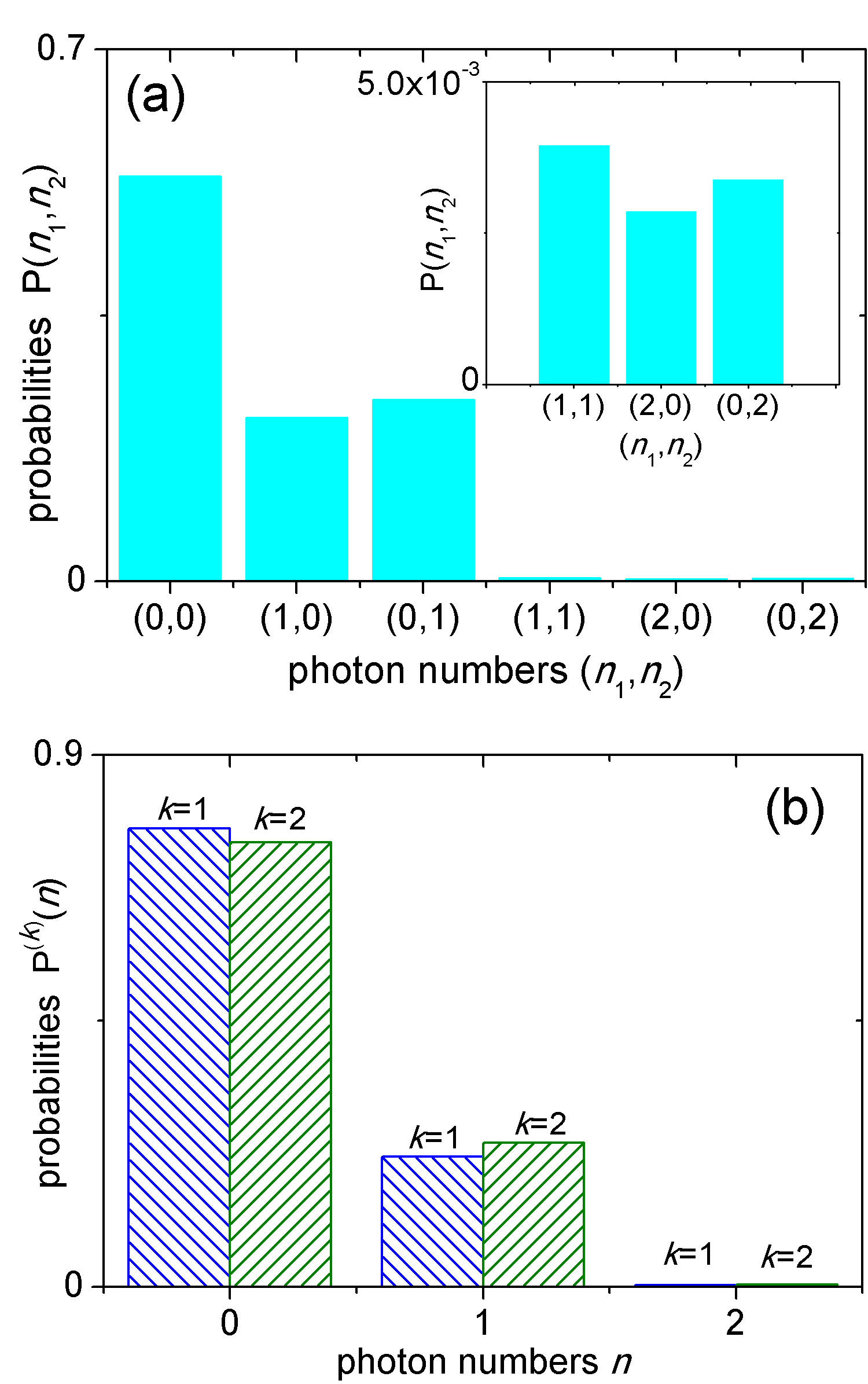} \caption{(Color
online) (a) Two-resonator photon-number probabilities
$P(n_{1},n_{2})$ and (b) photon-number probabilities $P^{(k)}$ of
the resonator $k=1$ and $k=2$ for the
output steady state (i.e., for $t\rightarrow\infty$) for the same
parameters as those in Fig.~\protect\ref{fig03}(b). It can be found
that the multi-photon states are hardly excited.} \label{fig04}
\end{figure}

With $\Gamma _{f}$ ($\Gamma $) denoting the pure dephasing (decay)
rate of the qubit, the evolution of the reduced density operator
$\rho(t)$ is governed by the standard Lindblad-Kossakowski master
equation,
\begin{eqnarray}
\frac{d\rho(t)}{dt} &=&-i[H,\rho(t)]+\Gamma D[\sigma _{-}]\rho(t)  \notag \\
&&+\frac{\Gamma _{f}}{2}D[\sigma _{z}]\rho(t)+\sum_{i=1,2}\gamma
_{i}D[a_{i}]\rho(t),  \label{eq21}
\end{eqnarray}
where the Lindblad superoperator $D$, acting on $\rho(t)$ with a
given collapse operator $B$, is defined by $D[B]\rho=B\rho B^{\dag
}-\frac{1}{2}(B^{\dag }B\rho-\rho B^{\dag }B)$. For simplicity, we
assume that all the parameters are dimensionless. By setting
$\beta =1,$ we choose the two resonators with the same frequency
$\omega _{i}=2500$ and the same coupling strength
$G_{i}=0.06\omega _{i}$. As a result, the effective quadratic
coupling strength is $\Theta =18$. We applied two coherent drives
with the unbalanced strengths $\epsilon _{1}=$ 0.95 and $\epsilon
_{2}=$ 1 for the two resonators. As a result, the two supermodes
$A_{+}$ and $A_{-}$ are driven resonantly with the strengths
$\epsilon _{+}=1.38$ and $\epsilon _{-}=-0.035$,
respectively (which can be calculated via the relations
shown in Table~\uppercase\expandafter{\romannumeral1}).

First, we consider the two supermodes are degenerate, i.e.,
$\Delta _{2}=0$. According to Eq.~(\ref{eq14}) we obtain the
direct coupling strength between the two resonators to be equal to
$g=6$. Defining the probabilities
\begin{align}
P(n_{1},n_{2})&=  \langle
n_{1},n_{2}|\rho(t)|n_{1},n_{2}\rangle,  \notag \\
P^{(k)}(n)&=  \langle n_{k}|\rho(t)|n_{k}\rangle,  \notag \\
P(\psi _{1+})&=\langle \psi _{1+}|\rho(t)|\psi_{1+}\rangle \notag
\end{align}
for the Fock states $|n_{1},n_{2}\rangle $, $|n_{k}\rangle$ (the
Fock states of the $k$th resonator), and the Bell-state $|\psi
_{1+}\rangle,$ we numerically simulate the original Hamiltonian
$H=H_{s}$ in Eq.~(\ref{eq4}) and the effective Hamiltonian
$H=H_{\text{eff}}$ in Eq.~(\ref{eq13}), respectively. The
time-dependent evolutions are plotted in Fig.~\ref{fig03}, where
subplot (a) [(b)] corresponds to the nondissipative (dissipative)
case.

It can be seen that the dynamical evolutions governed by
$H_{\text{eff}}$ (the curves marked with symbols) and $H_{s}$ (the
solid oscillating curves) match well each other in both
nondissipative and dissipative cases, indicating that the
approximations adopted for deriving the effective Hamiltonian are
valid. Since $\Theta \gg \epsilon _{+} $ and $\gamma _{1,2}\gg
\epsilon _{-}$, only the first excited state $|\psi _{1+}\rangle $
of the supermode $A_{+}$ can be excited effectively. Therefore,
the Hilbert space of two resonators is truncated into a
two-level system due the quadratic coupling. In
Fig.~\ref{fig03}(a), we find that the amplitudes of $P(0,0)$ and
$P(\psi _{1+})$ approximately exhibit qubit-like Rabi oscillations
without the consideration of any decay channel.

In Fig.~\ref{fig03}(b), we consider the dissipative case, and find
that the sum of $P(0,0)$ and $P(\psi _{1+})$ is almost equal to 1,
so the multiphoton probabilities $P(n_{1},n_{2})$ with
$n_{1}+n_{2}\geq 2$ are of extremely low amplitudes. In this case,
the two resonators behave as a single qubit. In Fig.~\ref{fig04},
we plot the probabilities for photon states $(n_{1},n_{2})$ and
the photon number distribution of each resonator for the original
Hamiltonian $H=H_{s} $ when $t\rightarrow\infty$. We find that,
for each resonator, only a single-photon state can be excited. For
the two resonators, the probabilities of multi-photon states are
all smaller than $5\times 10^{-3}$, while the states $P(0,0)$,
$P(0,1)$, and $P(1,0)$ are effectively occupied. This phenomenon
can be explained as single-photon two-resonator blockade; that is,
only one photon can be detected in these two resonators during two
zero-time-delay measurements.

Note that we describe the dissipative dynamics of our system by
the standard master equation under the Markov approximation and
assume weak couplings among all subsystems: the qubit, each
resonator, and the environment. To capture the non-Markovian
effects on the photon blockade, one can use, e.g., the effective
Keldysh action formalism~\cite{KamenevBook}, as recently applied
in a similar physical context in Ref.~\cite{Lemonde16}. Moreover,
to study photon blockade in our system in the ultrastrong or
deep-strong coupling regimes, the generalized master equation can
be applied within the general formalism of Breuer and Petruccione
(see sect. 3.3 in Ref.~\cite{BreuerBook}). This generalized master
equation was derived in detail for a circuit-QED system in
Ref.~\cite{Beaudoin11}. In that approach all subsystems (in our
case: the qubit and two resonators) would dissipate into a single
entangled channel. This is in contrast to the standard master
equation, as analyzed here, where we assume separable dissipation
channels for each subsystem.

In the following sections, we focus on the steady-state solutions
of the master equation, i.e., $\rho_{{\rm ss}}\equiv
\lim_{t\rightarrow \infty }\rho(t)$, by adopting the
time-independent Hamiltonian $H=H_{\text{eff}}$ and using a
shifted inverse power method implemented in
Ref.~\cite{Johansson13}.

\subsection{Phase-space description of photon blockade}

\begin{figure*}[tbp]
 \centering
 \includegraphics[height=6cm]{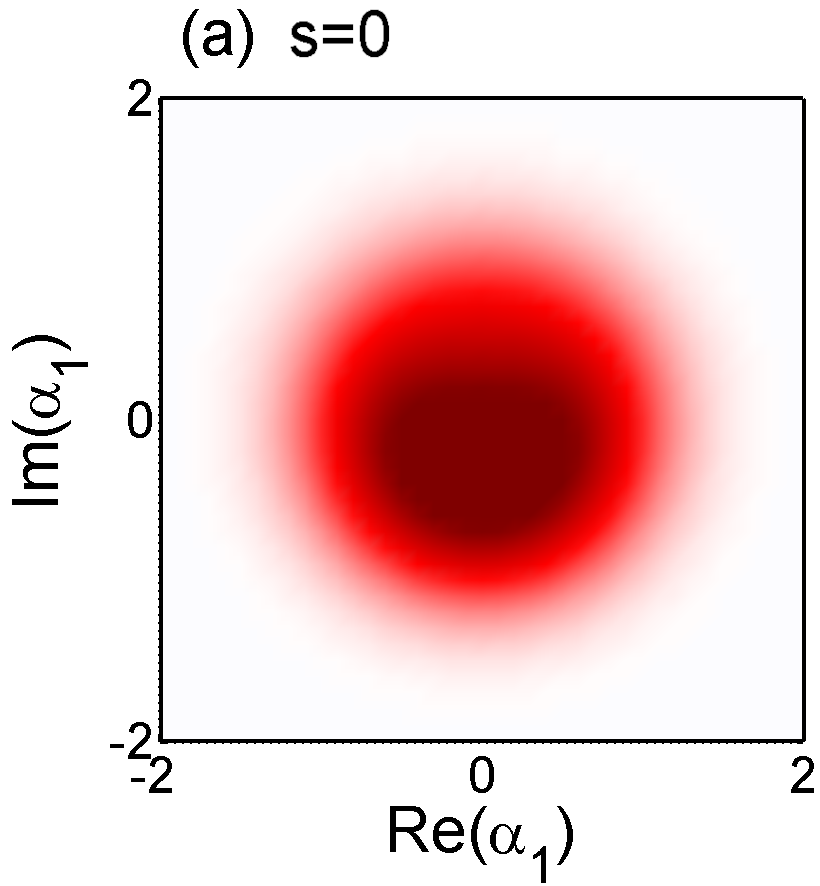}
 \includegraphics[height=6cm]{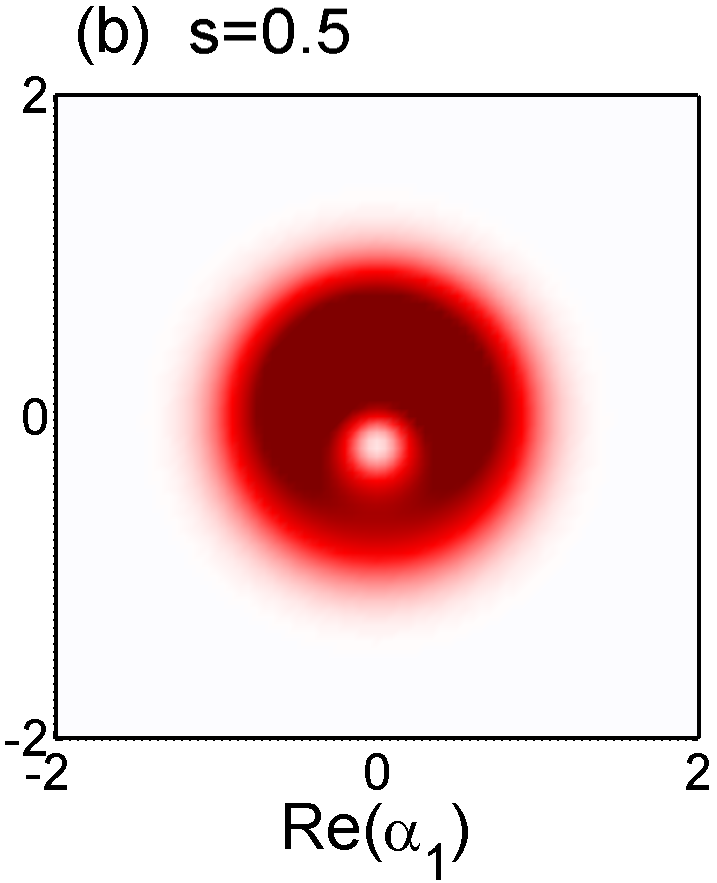}
 \includegraphics[height=6cm]{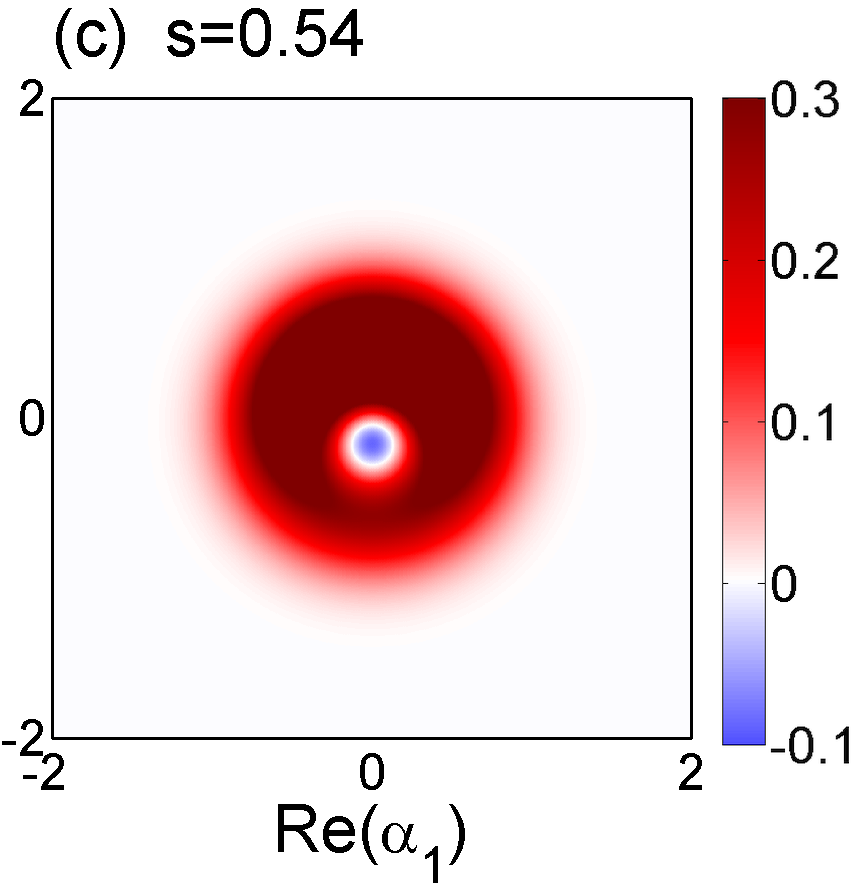}

\caption{(Color online) Single-resonator quasiprobability
distributions $W^{(s)}$ with parameter (a) $s=0$ (corresponding to
the Wigner function), (b) $s=1/2$,  and (c) $s=0.54$ for the
steady-state solutions $\rho^{(1)}_{{\rm ss}}={\rm
Tr}_{2}(\rho_{{\rm ss}})$ of the first resonator as a function of
its canonical position Re$(\alpha_1)$ and momentum Im$(\alpha_1)$.
The corresponding plots for the second resonator for
$\rho^{(2)}_{{\rm ss}}={\rm Tr}_{1}(\rho_{{\rm ss}})$ are very
similar to these and, thus, are not presented here. The other
parameters used here are the same as those in
Fig.~\protect\ref{fig03}(b).  The negativity of the QPD shown in
panel (c) clearly reveals the nonclassical character of the state
generated via photon blockade. We note that the parameter $s=0.54$
was chosen to be slightly larger than the nonclassical depth
$s_0=0.537$ of the state (or more precisely, of the corresponding
perfectly-truncated qubit state). Thus, the QPD shown in (c) is
non-positive, as indicated by the blue region.} \label{fig05}
\end{figure*}

To visualize the nonclassical properties of the fields generated
in our superconducting circuit, we apply the phase-space formalism
of Cahill and Glauber~\cite{Cahill69}. This formalism enables a
complete description of the dynamics of any quantum system in
terms of quasiprobability distributions (QPDs) and, thus, without
applying operators and their corresponding calculus (as in the
standard quantum-mechanical formalisms of, e.g., Schr\"odinger and
Heisenberg).

The Cahill-Glauber $s$-parametrized QPD, ${\cal W}^{(s)}(\alpha)$,
for $s\in[-1,1]$, of a given single-mode state $\rho$ can be
defined via its Fock state representation as
follows~\cite{Cahill69}:
\begin{equation}
  W^{(s)}(\alpha)=\sum_{k,l=0}^\infty\bra{k}\rho\ket{l}\bra{l}T^{(s)}(\alpha)\ket{k},
  \label{QPD1}
\end{equation}
given in terms of the operator $T^{(s)}(\alpha)$, which can be
defined via its Fock-state elements:
\begin{equation}
  \bra{l}T^{(s)}(\alpha)\ket{k} = c
  \sqrt{\frac{l!}{k!}}y^{k-l+1}z^l(\alpha^*)^{k-l}L^{k-l}_l(x_{\alpha}),
\label{QPD2}
\end{equation}
where $x_{\alpha}=4|\alpha|^2/(1-s^2)$, $y=2/(1-s)$,
$z=(s+1)/(s-1)$, $c=(1/\pi)\exp[-2|\alpha|^2/(1-s)]$, and
$L^{k-l}_l$ are the associated Laguerre
polynomials~\cite{SpanierBook}. The real and imaginary parts of
the QPD argument $\alpha$ are usually identified as the canonical
position and momentum, respectively. It is seen that this
$s$-parametrized QPD is a generalization of the Wigner $W$
function for $s=0$, the Husimi $Q$ function for $s=-1$, and the
Glauber-Sudarshan $P$ function in the limiting case for $s=1$.

The generalization of these single-mode QPDs for the multimode
case is straightforward. However, for brevity, we will not present
this generalization here, but will focus on the QPDs given by
Eq.~(\ref{QPD1}) for the single-mode (i.e., first-resonator)
reduced output states.

In Fig.~\ref{fig05} we plotted the $s$-parametrized QPDs for (a)
$s=0$ (which corresponds to the Wigner function), (b) $s=1/2$ and
(c) $s=0.54$ for a given choice of parameters of our system. These
plots are tomographic projections of the QPDs, where their
negative regions are marked in blue, as seen in
Fig.~\ref{fig05}(c) for some values of the canonical position
Re$(\alpha_1)$ and momentum Im$(\alpha_1)$ of the first resonator.
The negative regions of a given QPD reveal the nonclassical
character of the generated state. For a precise definition of
nonclassicality as well as its measures and witnesses see, e.g.,
Refs.~\cite{Miran10,Miran15noncl} and references therein. It is
seen that only the QPD shown in Fig.~\ref{fig05}(c) explicitly
shows the nonclassicality of the analyzed state. This
nonclassicality cannot be easily concluded by analyzing, e.g., the
nonnegative Wigner function in Fig.~\ref{fig05}(a).

The Cahill-Glauber formalism enables to define measures of
nonclassicality (or quantumness) of a quantum system. These
include the nonclassical depth $\tau$~\cite{Lee91} (for a recent
review see Ref.~\cite{Miran15noncl}). This measure can be defined
as the minimum amount of Gaussian noise (quantified by the
parameter $s$) required to destroy the nonclassicality or,
equivalently, to change the negative function $P\equiv{\cal
W}^{(1)}$ into a non-negative ${\cal W}^{(s_0)}$, i.e.:
\begin{equation}
{\cal W}^{(s_0) }\! ( \alpha) = \min_{s}c^{\prime}\! \int\!\!  \exp\left(
- \frac{2| \alpha - \beta|^2}{1-s} \right) {\cal W}^{(1)} ( \beta)
{\rm d}^2 \beta\ge0, \label{QPD4}
\end{equation}
where $s_0,s\in[-1,1)$ and $c^{\prime}=2/[\pi(1-s)]$. The Lee nonclassical
depth $\tau$ for a given state $\rho$ corresponds to this minimal
Cahill-Glauber parameter $s_0$ as follows
\begin{equation}
  \tau(\rho)=\frac12(1-s_0).
 \label{tau}
\end{equation}
Recently, it was shown that the nonclassical depth for a qubit
state, defined by the vacuum and single-photon states, is given
by~\cite{Miran15noncl}:
\begin{eqnarray}
    \tau(\rho) = \frac{\bra{1}\rho\ket{1}^2}{\bra{1}\rho\ket{1}-|\bra{0}\rho\ket{1}|^2}.
\label{noncl-depth}
\end{eqnarray}
Thus, if a perfect qubit state could be generated by photon
blockade, then its nonclassicality can be exactly given by
Eq.~(\ref{noncl-depth}). However, in our system we predicted the
generation of only effective non-perfect qubit states, which have
a minor contribution from the Fock states with a larger number of
photons. Specifically, the contribution of such terms is less than
$5\times10^{-3}$, as seen in the subset of Fig.~4(a). Such
imperfections of an effective qubit state result in its
nonclassical depth to be only approximately given by
Eq.~(\ref{noncl-depth}).

For the system parameters chosen in Fig.~\ref{fig05}, the
nonclassical depth is $\tau(\bar\rho^{(1)}_{{\rm ss}})=0.23$,
which corresponds to $s_0=0.537$, where $\bar\rho^{(1)}_{{\rm
ss}}$ is the single-resonator generated state $\rho^{(1)}_{{\rm
ss}}={\rm Tr}_{2}(\rho_{{\rm ss}})$, which is artificially truncated
to the qubit Hilbert space. The nonclassical depth for the state
$\rho^{(1)}_{{\rm ss}}$, which is calculated numerically with a
high precision in a higher-dimensional Hilbert space, is only
slightly larger than that obtained for the qubit truncated state
$\bar\rho^{(1)}_{{\rm ss}}$.

\subsection{Nonclassical photon-number correlations in photon blockade}
\begin{figure}[tbp]
\centering \includegraphics[width=8.8cm]{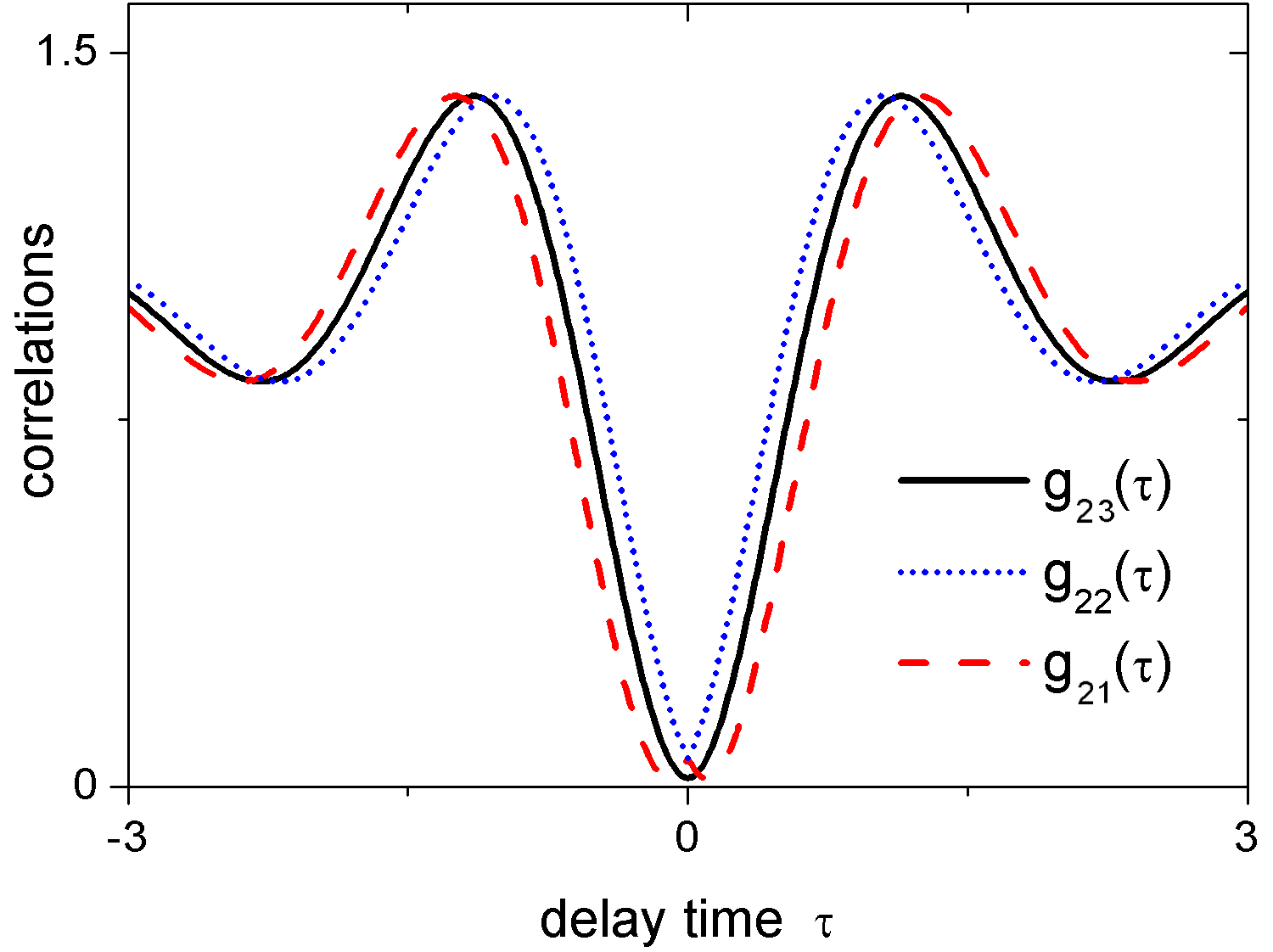} \caption{(Color
online) The two-time second-order correlation functions
$g_{21}(\tau)$ (red dashed curve), $g_{22}(\tau)$ (blue dot curve)
and $g_{23}(\tau)$ (black solid curve) as functions of the delay
time $\tau$ for ports 1, 2 and 3. All these three second-order
correlation functions are much smaller than 1, and show dips at
zero-time delay $\protect\tau =0$. These dips reveal strong
sub-Poissonian photon number statistics, since
$g_{2i}(\tau=0)\approx 0$, while the increase of $g_{2i}(\tau)$
with increasing $\tau$ from  $\tau=0$ reveal photon
antibunching. The parameters used here are the same as those in
Fig.~\protect\ref{fig03}(b).} \label{fig06}
\end{figure}

Here we analyze nonclassical photon-number correlations of the
stationary output fields generated in our superconducting system.
We will show that the output signals in all the three ports
can exhibit both sub-Poissonian photon-number statistics and
photon antibunching under appropriate conditions.

Let us define the time-delay second-order
correlation function of the output field of the steady state as
\begin{equation}
g_{2i}(\tau)=\lim_{t\rightarrow \infty }\frac{\langle c_{i}^{\dag
}(t)c_{i}^{\dag }(t+\tau )c_{i}(t+\tau )c_{i}(t)\rangle }{\langle
c_{i}^{\dag }(t)c_{i}(t)\rangle \langle c_{i}^{\dag }(t+\tau
)c_{i}(t+\tau )\rangle },  \label{eq19}
\end{equation}where $\tau $ is the time-delay between two measurements. At $\tau
=0$, the second-order correlation function of the three output
ports can be expressed as
\begin{subequations}
\begin{align}
g_{21}(0)& =N_{1}^{-2}\langle a_{1}^{\dag }a_{1}^{\dag
}a_{1}a_{1}\rangle,
\label{eq20a} \\
g_{22}(0)& =N_{2}^{-2}\langle a_{2}^{\dag }a_{2}^{\dag
}a_{2}a_{2}\rangle,
\label{eq20b} \\
g_{23}(0)& = \frac{1}{9N_{3}^{2}}\sum_{j,k,m,l=1,2}\sqrt{\gamma
_{i}\gamma _{k}\gamma _{m}\gamma _{l}}\langle
a_{j}^{\dag}a_{k}^{\dag }a_{m}a_{l}\rangle.  \label{eq20c}
\end{align}
\end{subequations}
Several previous studies on photon (phonon) blockade equate the
concepts of photon antibunching and sub-Poissonian photon-number
statistics. However, we treat these two effects distinctly according
to their standard definitions. Specifically, for stationary
fields, photon antibunching (bunching) means that $g_{2i}(\tau )>$
$g_{2i}(0)$ [$g_{2i}(\tau )$ $<g_{2i}(0)$], that is, a local
minimum (maximum) around the zero-time
delay~\cite{MandelBook,Miran10}. The sub-Poissonian
(super-Poissonian) photon statistics only indicate that
$g_{2i}(0)<1$ [$g_{2i}(0)>1$]. Thus, sub-Poissonian statistics
does not imply photon antibunching and vice versa~\cite{Zou90}.
Note that both photon antibunching and sub-Poissonian statistics
are key features for an ideal single-photon source. We will show
that both of these two purely nonclassical effects can be observed
in our proposal.

As discussed in Sec.~III.A, given that only states $|00\rangle $,
$|10\rangle,$ and $|01\rangle $ are of large probabilities, while
the two-photon states $|20\rangle $, $|02\rangle,$ and $|11\rangle
$ are of extremely low probabilities, we will observe
single-photon blockade in two resonators: A single photon in one
resonator cannot only blockade the second photon in this
resonator, but can also blockade another photon from being excited
in another resonator. Consequently, once the photon escapes from
the two resonators can the system be reexcited. As a result, the
photon distribution from ports 1 and 2 are both sub-Poissonian,
and the cross-correlation between two resonators displays the
anticorrelation. Moreover, in the following sections, we will show
that the field from port 3 also exhibits both sub-Poissonian
statistics and antibunching. Thus, a single photon can be emitted
from ports 1 and 2, or alternatively from port 3.

In Fig.~\ref{fig06}, the time-delay second-order correlation
functions $g_{2i}(\tau)$ of the steady state of the output port
$i$ are plotted, from which we find that $g_{2i}(\tau )\ll 1$ and
all show dips at $\tau =0,$ indicating that the output microwave
fields from the three ports exhibit both sub-Poissonian
photon-number statistics and photon antibunching.

\begin{figure*}[tbp]
\centering
\includegraphics[width=1.00\linewidth]{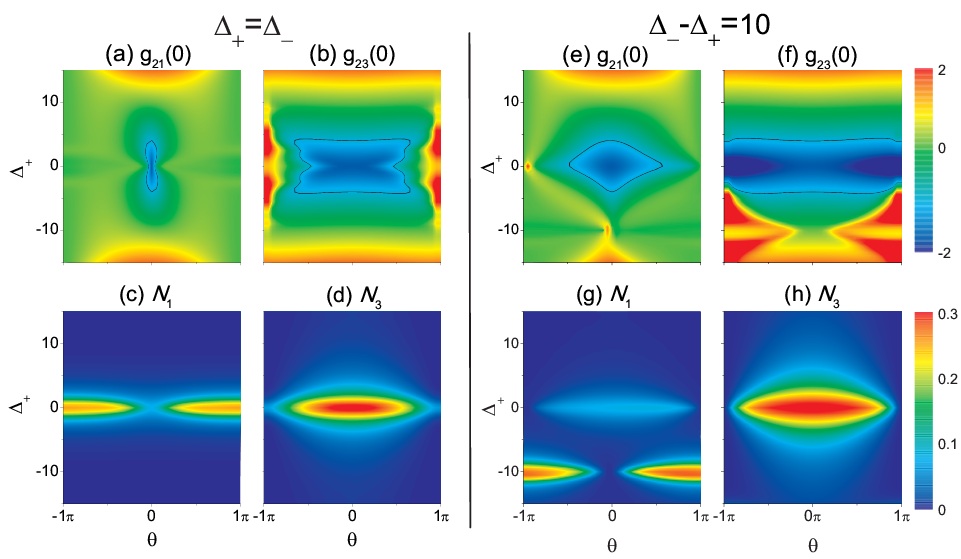}
\caption{(Color online) The average photon escape rate $N_{1}$
($N_{3}$) and the second-order correlation function $g_{21}(0)$
[$g_{23}(0)$] from port 1 (port 3) versus the drive detuning
$\Delta _{+}$ and the phase difference $\protect\theta $. The left
panel corresponds to the degenerate case ($\Delta _{2}=0$) while
the right panel shows the nondegenerate case ($\Delta _{2}=10$).
Experiments could adjust the coupling capacity $C$ between the
two resonators, to change the frequency separation between the two
supermodes. In the plot of the second-order correlation function
$g_{2i}(0)$, the solid black closed loops in (a, b, e, f)
correspond to $\log _{10}[g_{2i}(0)]=-1$ and the points inside the
loops indicate that the output fields exhibit a strong
sub-Poissonian character. It can be found that, in both degenerate
and nondegenerate cases, $g_{21}(0)$ and $g_{23}(0)$ can display
dips around $\Delta _{+}=0$ and $\protect\theta =0$. The other
parameters used here are the same as those in
Fig.~\protect\ref{fig03}(b).} \label{fig07}
\end{figure*}

In Fig.~\ref{fig07}, we plot the computed values of $\log
_{10}[g_{2i}(0)]$ and $N_{i}$ changing with $\theta $ and $\Delta
_{+}.$ Here we assume that the drives for the two resonators are
of the same strength, while the phase difference between the two
microwave drives is $\theta $, i.e, $\epsilon _{1}=\epsilon
_{2}^{\ast }=\exp (i\theta /2$), and the corresponding driving
strengths for the supermodes $A_{+}$ and $A_{-}$ are $\epsilon
_{+}=2\cos (\theta /2)$ and $\epsilon _{-}=2i\sin (\theta /2)$,
respectively. It is obvious that $\epsilon_{+}$ ($\epsilon _{-}$) decreases (increases)
with increasing $|\theta |$ in the regime $[0,\pi ].$ Since the
resonators 1 and 2 are identical, the modes $a_{1}$ and $a_{2}$
share the same dynamics and, thus, we only plot $g_{21}(0)$ and
$N_{1}$. In particular, due to $\gamma _{1}/\gamma _{2}=\beta =1,$
for port 3 we have
\begin{gather*}
N_{3}\propto \langle A_{+}^{\dag }A_{+}\rangle , \\
g_{23}(0)\propto \langle A_{+}^{\dag }A_{+}^{\dag
}A_{+}A_{+}\rangle /\langle A_{+}^{\dag }A_{+}\rangle ^{2},
\end{gather*}so the photon statistics of the output field from port 3 is
determined only by the properties of the supermode $A_{+}$ under
these conditions.

In the left panel of Fig.~\ref{fig07}, we consider that the two
supermodes are degenerate with $\Delta _{2}=0$. Around $\Delta
_{+}=0$ (i.e., the drive for the supermode $A_{+}$ is resonant),
both $g_{21}(0)$ and $g_{23}(0)$ show a dip at $\theta =0$.
However, with increasing $|\theta|$, the driving strength
$\epsilon _{-}$ for the supermode $A_{-}$ goes up,
leading to its eigenstates $|\psi _{i-}\rangle $ being effectively
excited. It can easily be verified that mode $a_{1}$ satisfies
\begin{subequations}
\begin{gather}
\langle \psi _{i-}|a_{1}^{\dag }a_{1}|\psi _{i-}\rangle \neq 0,
\label{eq22a} \\
\langle \psi _{j-}|a_{1}^{\dag }a_{1}^{\dag }a_{1}a_{1}|\psi
_{j-}\rangle \neq 0,  \label{eq22b}
\end{gather}
\end{subequations}
while for supermode $A_{+}$:
\begin{subequations}
\begin{gather}
\langle \psi _{i-}|A_{+}^{\dag }A_{+}|\psi _{i-}\rangle =0,  \label{eq23a} \\
\langle \psi _{j-}|A_{+}^{\dag }A_{+}^{\dag }A_{+}A_{+}|\psi
_{j-}\rangle =0, \label{eq23b}
\end{gather}where $i\geq 1$ and $j\geq 2$. Thus, the eigenstates of the supermode
$A_{-}$ being effectively excited lead to the increase of both
output photon number $N_{1}$ and second-order correlation function
$g_{21}(0)$. However, their contributions to $N_{3}$ and
$g_{23}(0)$ vanish according to Eqs.~(\ref{eq23a}) and
(\ref{eq23b}). In Figs.~\ref{fig07}(a) and \ref{fig07}(b), it can be
found that, compared with $g_{23}(0),$ $g_{21}(0)$ is much more
sensitive to the changes of $\theta $: even though $\theta $ is a
slightly bias from $0$, the photon statistics of the output field
from port 1 will not be sub-Poissonian any more. In
Fig.~\ref{fig07}(c) we find that, around $\Delta _{+}=0$, the
average photon number $N_{1}$ from port 1 increases with $|\theta
|$. At $\theta =\pm \pi $, the drive strength for the supermode
$A_{-}$ and the photon number in resonator 1 both reach their
maxima. However, the field from port 1 is not sub-Poissonian any
more. The photon output $N_{3}$ from outport 3 vanishes at $\theta
=\pm \pi $, as shown in Fig.~\ref{fig07}(d), for two reasons:
first, the drive strength $\epsilon _{+}$ for the supermode
$A_{+}$ decreases to zero; second, there is no contribution from
the eigenstates $|\psi _{j-}\rangle $ of the supermode $A_{-}$.

In the right panel of Fig.~\ref{fig07}, we plot the non-degenerate
case with $\Delta _{-}-\Delta _{+}=10.$ By comparing with the
degenerate case, we find that both $g_{21}(0)$ and $g_{23}(0)$
display the sub-Poissonian behavior in a wider range of $\theta$.
In this case, despite of increasing $\theta$, the driving strength
$\epsilon _{-}$ for the supermode $A_{-} $ is still far
off-resonance around $\Delta _{+}=0$, as shown in Fig.~\ref{fig02}.
Therefore, the states $|\psi _{i-}\rangle $ cannot be effectively
excited, so their contributions for the mode $a_{1}$ are
negligible. Only the driving $\epsilon _{+}$ for the supermode
$A_{+}$ affects $N_{1}$ and $g_{21}(0)$. Due to the nonlinear
coupling between the supermode $A_{+}$ and the qubit, only the
state $|\psi _{1+}\rangle $ can be excited effectively. Compared
with the degenerate case in Fig.~\ref{fig07}(a), $g_{21}(0)$ in
Fig.~\ref{fig07}(e) is less sensitive to the phase difference
$\theta$. In Fig.~\ref{fig07}(g), we find that, around $\Delta
_{+}=-10$, the photon number $N_{1}$ from port 1 is very large,
owing to the resonant driving of the supermode $A_{-}$. Since
there is no nonlinear coupling between the supermode $A_{-}$ and
the qubit, multi-photon states for the supermode $A_{-}$ are
excited. Although the output photon number $N_{1}$ is large, the
second-order correlation function $g_{21}(0) $ is not
sub-Poissonian.

Last, we want to discuss another interesting phenomenon.
Specifically, if the direct coupling $g$ between the two
resonators vanishes (i.e., the capacitor $C$ is removed), the two
supermodes $A_{+}$ and $A_{-}$ are still nondegenerate, and the
frequency difference is only determined by the dispersive coupling
strength, as shown in Eq.~(\ref{eq14}), i.e, $\Delta
_{2}=4G_{x}^{2}/\left( 3\Omega _{+}\right) $. In this case, even
when only one resonator is under a resonantly-coherent drive (for
example, $\epsilon _{1}=1$ and $\epsilon _{2}=0$), the phenomenon,
that single photon outputs from ports 1, 2, and 3, still exists
under the condition
\end{subequations}
\begin{equation}
\Delta _{2}=4G_{x}^{2}/\left( 3\Omega _{+}\right) \gg \epsilon_{-},
\end{equation}which can easily be realized in experiments. Thus, by employing
only one coherent drive, and one auxiliary qubit without any direct
coupling between two resonators, the single-photon outputs also
exist in all the three output channels.

\subsection{Entanglement relation between the two cavities}
\begin{figure}[tbp]
\centering \centering \includegraphics[width=8.8cm]{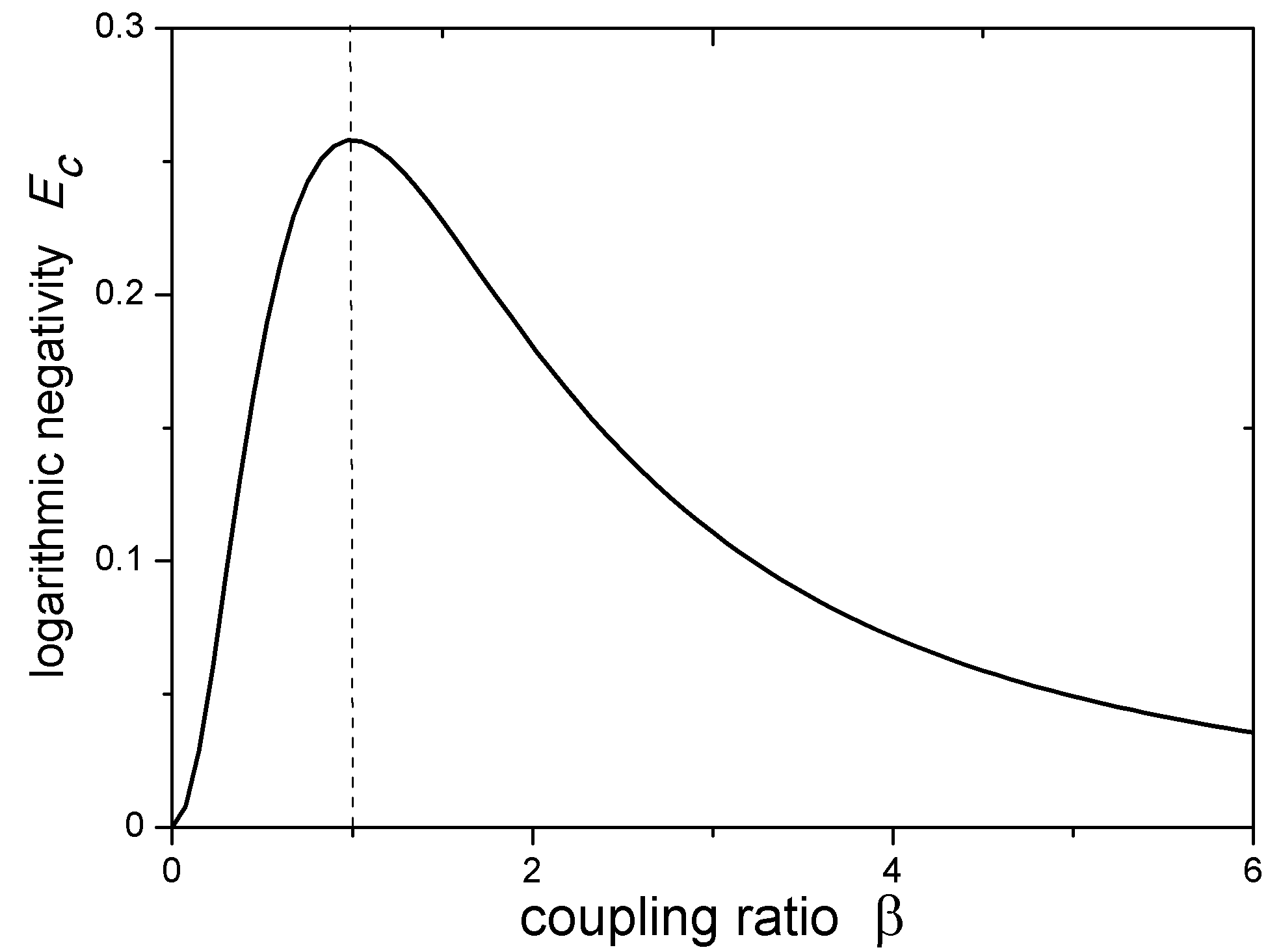}
\caption{The logarithmic negativity $E_{c}$ versus the coupling
ratio $\protect\beta $. Here we fix $\protect\theta =0$ and
$\Delta _{+}=0$. Other parameters are the same as those of the
nondegenerate case in Fig.~\ref{fig07}. The vertical dashed line is
at $\protect\beta =1$.} \label{fig08}
\end{figure}

To analyze the entanglement between the resonators, we use the
logarithmic negativity to measure the entanglement, which is given
by~\cite{Horodecki09}
\begin{equation}
 E_{c}=\log _{2}\left[2 N_{E}(\rho_{12})+1\right], \label{eq25}
\end{equation}where the negativity $N_{E}({\rho}_{12})$ quantifies
the entanglement of the two-resonator steady state $\rho_{12}$,
which can be expressed as
\begin{equation}
N_{E}({\rho}_{12})=\frac{||{\rho}_{12}^{T_{1}}||-1}{2}.
\label{eq26}
\end{equation}Here $T_{1}$ denotes the partial transpose of the density
matrix ${\rho}_{12}$ with respect to the resonator 1, and
$||{\rho}_{12}^{T_{1}}||$ is the trace norm of
${\rho}_{12}^{T_{1}}.$ The logarithmic negativity $E_{c}$ is
an entanglement monotone, which can be used for quantifying the
entanglement between the two resonators (i.e., the entanglement
between signals from ports 1 and 2). In Fig.~\ref{fig08}, we adopt
the parameters with $\theta =0$ and $\Delta _{2}=10$, and plot the
dependence of $E_{c}$ on the ratio $\beta $. In this case, only
the first excited state $|\psi _{1+}\rangle $ of the supermode
$A_{+}$ can be effectively driven. Note that $|\psi _{1+}\rangle$
is a maximally-entangled state (i.e., the `triplet' state)
when $\beta =1$. As shown in Fig.~\ref{fig08}, we find that the
output fields from ports 1 and 2 are entangled, and the
logarithmic negativity $E_{c}$ reaches its maximum value when
$\beta =1.$ We find that, the entanglement between fields from
these two ports has a close relation with the single-photon
blockade effects, which originates from optical state truncation
(or the nonlinear quantum scissors). That is, the states of the
two cavities are truncated to a qubit, with the single-photon Bell
`triplet' state $|\psi_{1+}\rangle $ being the first-excited
state.
\section{Discussion and conclusions}
In this paper we demonstrated that it is possible to achieve
single-photon outputs in a circuit-QED system based on \emph{both}
longitudinal and transverse couplings. We obtained the effective
Hamiltonians and the rates for multi-photon processes, and found
that the effective nonlinear coupling between one of the
supermodes and the qubit can lead to photon blockade effects.

We note the multi-photon processes can also be induced in the
hybrid superconducting system with only longitudinal coupling,
which has been shown in our previous study~\cite{Wangx16}. In this
work, we found that the second-order nonlinearity can be about one order of magnitude
stronger. Moreover, the drive for the qubit is not needed in the
present case.

We have analyzed photon blockade in phase space by applying the
Cahill-Glauber $s$-parametrized QPDs. This approach enabled us to
show not only the nonclassical character of the states generated
via photon blockade, but also to determine the degree of
nonclassicality of the states, using the Lee nonclassical depth.

Moreover, we considered two different output channels for the
fields: those from the individual resonators and the joint
channels of both resonators. It was found that all the three
output fields display photon antibunching and sub-Poissonian
photon-number distribution. Thus, our proposal can be used to work
as multi-output single microwave photon devices. Afterwards, by
analyzing the steady-state solutions, we discussed the degenerate
and nondegenerate cases of the two supermodes. In the degenerate
case, the second-order correlation function $g_{21}(0)$ of port 1
is much more sensitive to the increase of the drive strength
$\epsilon _{-}$ for the supermode $A_{-}$ than in the
nondegenerate case. For the joint output 3, due to no contribution
from the eigenstates of the supermode $A_{-}$, $g_{23}(0)$ is more
robust against the increase of $\epsilon _{-}$ than that of
$g_{21}(0)$ in both degenerate and nondegenerate cases. We also
found that the state truncation of two-resonator modes will lead
to the entanglement between two resonators.

Compared with the dispersive and resonant microwave-photon
blockade known from previous studies, our proposal has the
following two advantages: first, the excited state of the
system still retains a photonic nature (i.e., this is a pure
single-photon Fock state rather than a polariton state); second,
the strong nonlinearity makes it possible to increase the
single-photon output rate.

It should be stressed that, to obtain multi-output channels, we
consider two resonators in this paper, but these results can also
be applied to the simple \emph{single-resonator case}. We believe that
our proposal can be helpful in designing single-photon sources in
the microwave regime.

\section*{Acknowledgments}

The authors acknowledge fruitful discussions with Anton F. Kockum,
Wei Qin, Salvatore Savasta, Chui-Ping Yang, and Zhi-Rong Zhong. XW
is supported by the China Scholarship Council (Grant No.
201506280142). AM and FN acknowledge the support of a grant from
the John Templeton Foundation. FN was also partially supported by
the RIKEN iTHES Project, the MURI Center for Dynamic
Magneto-Optics via the AFOSR award number FA9550-14-1-0040, the
IMPACT program of JST, CREST, and a grant from the Japan Society for the Promotion of Science (KAKENHI).

\end{document}